\documentclass[twocolumn]{revtex4} 

\usepackage{amsmath}
\usepackage{graphicx}
\usepackage{SIunits}

\begin{document}

\title{Microwave sidebands for atomic physics experiments by period one oscillation in optically injected diode lasers}

\author{C. I. Laidler and S. Eriksson}

\affiliation{Department of Physics, Swansea University, Singleton Park, Swansea SA2 8PP, United Kingdom
}

\begin{abstract}
We show that nonlinear dynamics in diode lasers with optical injection leads to  
frequency tunable microwave sidebands which are suitable for atomic physics experiments. 
We demonstrate the applicability of the sidebands in an experiment where rubidium atoms are magneto-optically trapped with both the trap and the re-pump optical frequencies derived from one optically injected laser.  We find linewidth narrowing in the optical spectrum of the injected laser at both the injection frequency and the sideband frequency. With strong optical injection which leads to frequency locking we find a complete linewidth transfer from the master to the slave. Further applications are discussed.
\end{abstract}

\maketitle

High power and narrowband emission are typical prerequisites of continuous wave lasers. Applications in atomic physics such as laser cooling and trapping~\cite{Metcalf1999} often require two or more lasers with these qualities. Laser cooling of carefully selected molecules~\cite{Schuman2010} is possible, but  several lasers are necessary to close the transitions for efficient cooling.  A cost-effective way to produce high power narrowband light is to injection lock a diode slave laser with high power but often  poor spectral properties to a narrow linewidth master laser~\cite{Schunemann1998}. This technique utilises one of the many nonlinear dynamical states of a diode laser with external optical injection~\cite{Wieczorek2005}. Recently, another dynamical state in the slave laser,  the period one (P1) oscillation limit cycle of undamped relaxation oscillations, has received increasing attention~\cite{Chan2004,Chan:07,chan2010}. Remarkably, the frequency of the P1 oscillation can far exceed the modulation bandwidth of free running diode lasers. As a consequence, optical injection  can be used to produce fast modulations in the microwave range of the spectrum even at frequencies not typically covered by standard external modulators. 

In this letter we report on the general applicability of P1 oscillation dynamics to experiments in atomic physics. We show that by carefully controlling the injection we can produce tunable sidebands. We demonstrate the sideband stability by creating a magneto-optical trap (MOT) with light solely from the optically injected slave laser. We show that the frequencies needed for a MOT can be produced both with weak injection which does not lead to frequency locking of the slave and strong injection with frequency locking on the master laser frequency. We measure the optical spectrum of the slave laser with a heterodyne technique and find that  the spectral features at both the injection frequency and at the sidebands frequency are narrower than the free running slave laser. In the frequency locked case we find a complete linewidth transfer from master to slave. The sidebands in the optically injected slave laser can extend to several tens of GHz ~\cite{Chan2004}. We discuss applications of the sideband technique beyond laser cooling of alkalis.

\begin{figure}[htb]
\begin{center}
\includegraphics[width=0.9\linewidth]{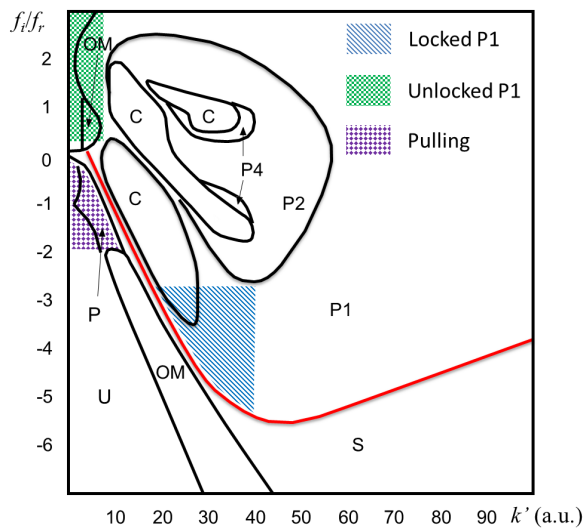}
\end{center}
\caption{Map of the most prominent dynamical regions in an edge emitting diode laser. The patterned regions (further explained in the main text) show roughly where we explored the P1 dynamics which we used in this work. Some details have been omitted for clarity when adapting the diagram from \cite{eriksson2002}. \label{fig:map} 
}
\end{figure}

In an edge emitting diode laser typical values of the laser gain and cavity decay rates lead to relaxation-oscillations in the population inversion and the intensity. The optical spectrum of a typical single mode diode laser consists of a main peak centred on the laser oscillation frequency and very weak sidebands separated by the relaxation oscillation frequency $f_r$ from the peak. The modulation bandwidth of the free running diode laser is associated with $f_r$ which above threshold is proportional to the square root of the injection current. When optical injection is introduced the beat note at the master-slave laser frequency difference, $f_i=f_{ml}-f_{sl}$, drives the relaxation oscillation with a strength which depends on the injected power relative to the free running slave power, $k = P_{inj}/P$. Suitable combinations of the experimentally adjustable $f_i$ and $k$ can lead to spectacular phenomena in the slave output such as deterministic chaos~\cite{Simpson1997} with periodic windows~\cite{Eriksson2001}. The asymmetry of the semiconductor gain along the frequency axis leads to a coupling between amplitude and phase fluctuations with a coupling strength which is measured by the linewidth-enhancement factor $\alpha$. One consequence of this coupling is that when the population inversion is modified by optical injection, the frequency of the laser mode shifts towards lower frequency with increased injection power. In practice this means that the sign of $f_i$ is relevant and for typical diode lasers the  injection locking range in the ($f_i$, $k$)-plane opens up asymmetrically around zero $f_i$ towards negative $f_i$. The coupling also leads to a complicated arrangement of  bifurcations in the ($f_i$, $k$)-plane~\cite{Wieczorek1999}. 
However, the dynamical states have  been experimentally mapped out~\cite{Simpson1997,eriksson2002,eriksson2002b} and the regions in the parameter space which produce unstable oscillations can be avoided if the application so demands. It has also been shown that a rate-equation model captures the essential features of the nonlinear dynamics observed in experiments, and the dynamical complexity is well understood~\cite{Krauskopf2000, Wieczorek2005}. Both experiment and theory show that  the arrangement of dynamical regions in the ($f_i$, $k$)-plane depends strongly on $\alpha$. However, the value of $\alpha$ is similar in most edge-emitting lasers and the overall shape of the regions within the map remains qualitatively similar from laser to laser. Fig.~\ref{fig:map} shows a map of the main dynamical regions of an edge emitting diode laser in the ($f_i$, $k$)-plane~\cite{eriksson2002}. 
Most applications utilise the stable injection locked region labelled 'S' in Fig.~\ref{fig:map} where the slave laser is oscillating with a single mode output at the master laser frequency. In this work we focus on the region in the ($f_i$, $k$) plane where the motion of the laser electric field envelope and population inversion forms a closed trajectory in phase space within one period. This motion is referred to as P1 oscillation and it occurs when $f_i$ is positive with respect to the frequency boundary formed by the Hopf-bifurcation line. In Fig~\ref{fig:map} the Hopf-bifurcation line forms a  boundary between the stable injection locking region labelled 'S' , and the P1 region. We avoid the period doubling regions (P2 and P4) and chaos (C).

Our master laser power was insufficient to explore regions corresponding to  $k'> 40$  in Fig.~\ref{fig:map} for the injection current used in the trapping experiments. We mapped out a smaller portion of the dynamics in this work in detail and we observe small qualitative changes. Most notably we observe an absence of the signatures of chaos in the lowermost island and an overall reduction of the size of the unstable regions. From this we expect that our $\alpha$ factor is slightly smaller than in ref.\cite{eriksson2002}. Jumps to other modes (OM) also occur in our laser for low $k$ before the frequency locks to the master laser, but in somewhat different locations. Nevertheless we found that overall Fig.~\ref{fig:map} served as a good guide to the location of the dynamics.  We note that the P1 region extends much further towards positive $f_i$ than this diagram shows.

\begin{figure}[htb]
\centerline{\includegraphics[width=0.8\linewidth]{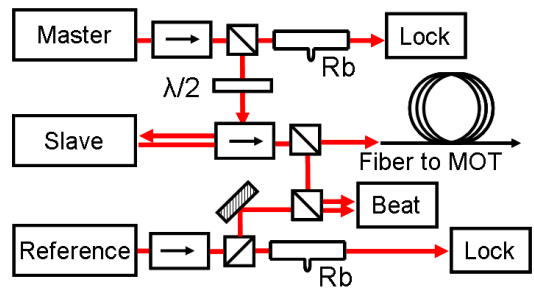}}
\caption{Diagram of the experiment. The details of the
Doppler-free spectrometers have been omitted for clarity\label{fig:experiment}.}
\end{figure}

The experiment is arranged as shown in Fig.~\ref{fig:experiment}. 
Light from a home-made extended cavity diode laser is injected via the output polariser of a Faraday isolator into a collimated and temperature stabilized Sharp GH0781JA2C edge emitting diode laser. We control the injection intensity with a $\lambda$/2-plate. A small fraction of the output from the slave laser is sampled and combined with a reference laser using beam splitters before launching into a single mode polarisation preserving optical fibre which guides the light into the MOT. The fibre delivery ensures that the spatial profile of the MOT light is independent of the source which allows for a straightforward comparison of MOTs made with different light sources. The combined slave and reference laser beams interfere on a low bandwidth ($<$\unit{200}{\mega\hertz}) photodetector. We either keep the  reference laser frequency fixed on a rubidium hyperfine transition so that the heterodyne signal can be used directly to study linewidths, or we sweep the reference laser fast over its mode-hop free tuning range of \unit{8}{\giga\hertz}. In this case the heterodyne signal is low pass filtered to produce the overall slave laser spectrum. A small fraction of both the master and reference laser intensity is directed to individual rubidium vapour cell polarisation spectrometers~\cite{Wieman1976, Pearman2002} for laser frequency stabilization. We found it necessary to separately isolate the master laser spectrometer from the slave laser (not shown in Fig.~\ref{fig:experiment}). The master and reference lasers can be tuned to any Doppler-free feature of the entire absorption spectrum of the D2 manifold of rubidium. The slave laser is chosen so that the frequency of a longitudinal mode is near enough to the master laser to produce the desired $f_i$. Our slave laser can be tuned several \unit{\giga\hertz} which helps us explore its dynamics widely. For a fixed application using strong injection this tuning requirement is less stringent because the strong injection will  force the slave to operate with power predominantly at the injection frequency. However, for experiments requiring weak injection the free running laser must operate with the dominant longitudinal mode at $f_i$ from the master frequency. The MOT is formed by retro-reflecting three beams aligned to intersect at the center of a quadrupole magnetic field with a \unit{0.11}{\tesla\per\meter} axial field gradient. The beam was collimated with a Thorlabs F810APC-780 fibre collimation package which has a specified $e^{-2}$ output beam diameter of \unit{7.5}{\milli\meter}. We  make MOTs with total power in the beams from about \unit{5}{\milli\watt} to a maximum of  \unit{10}{\milli\watt}. We dispense rubidium atoms into the vacuum chamber using a  getter source pulsed at 8A for 10 seconds and we measure the fluorescence from the trapped atoms.

We find two P1 oscillation regions in the $(f_i,k)$ plane which produce tunable sidebands in the \unit{\giga\hertz} range. We classify these two dynamical states of the slave laser according to whether the slave laser frequency is locked to the master laser frequency or not. We surmise that the phase difference between the laser input and output is bounded for the injection locked laser but unbounded (i.e. running)~\cite{Wieczorek2005} when the frequency is unlocked. However, our linewidth purity transfer measurements which we detail below indicate that even when the slave laser frequency is unlocked the master laser provides at least some phase reference. When $f_i > 0$ and $k$ is too weak to injection lock the frequency of the slave laser to the master frequency an unlocked  P1 oscillation results (checkered region in Fig~\ref{fig:map}). The injection modifies the slave laser gain and as a result of the amplitude-phase coupling the slave laser frequency is shifted from the free running frequency towards lower absolute frequency by an amount which depends on $k$. The slave laser spectrum consists of a weaker peak at the master laser optical frequency and separated by the P1 oscillation frequency, $f_s$, a stronger peak at the shifted free running slave optical frequency. An additional much smaller sideband appears at frequency $f_s$ below the slave laser peak. When $f_i < 0$, but still positive with respect to the Hopf-bifurcation line~\cite{Wieczorek1999} the slave laser is frequency locked to the master and we find a frequency locked P1 oscillation (hatched region in Fig.~\ref{fig:map}).  The slave laser spectrum is  qualitatively indistinguishable from the unlocked case, but now the master and slave laser frequencies coincide. A feature with practical consequences is that there is now also a coherent  phase relationship between the master and slave lasers. Both the locked and unlocked P1 oscillation frequencies are tunable from about \unit{2}{\giga\hertz} to at least 21.2 {\giga\hertz} by changing the injection parameters. The larger sideband frequencies were observed by shifting the central frequency of the reference laser sweep away from the rubidium resonances whilst measuring the frequency on a wavemeter. The sideband to carrier intensity ratio also depends on the injection parameters and a suitable combination of $k$ and $f_i$ which produces a desired ratio can be found~\cite{Chan:07}. 

In a third region of the $(f_i,k)$ plane where $f_i<0$ and $k$ is too weak to produce injection locking we find that the frequency of the free running slave mode is pulled towards the master laser frequency as $k$ is increased (diamonds in Fig.~\ref{fig:map}). A weaker peak in the slave laser spectrum appears at the master laser frequency $f_s$ away from the free running mode and further peaks appear at the harmonics of $f_s$. As $k$ is further increased the slave eventually becomes frequency locked with single frequency output. We believe this frequency pulling dynamics, which was also observed in~\cite{eriksson2002}, was used in ref.~\cite{Moon1996} to make both frequencies of light for MOTs. In our experiment the achieved frequency interval was always below the hyperfine splitting in rubidium, except for very low injection powers which could only be achieved by deliberate misalignment of the injection beam. Therefore, we did not have sufficient control to explore this oscillation further. We turn next to detailed measurements of the P1 oscillation dynamics with larger sideband frequencies.

\begin{figure}[htb]
\centerline{\includegraphics[width=\linewidth]{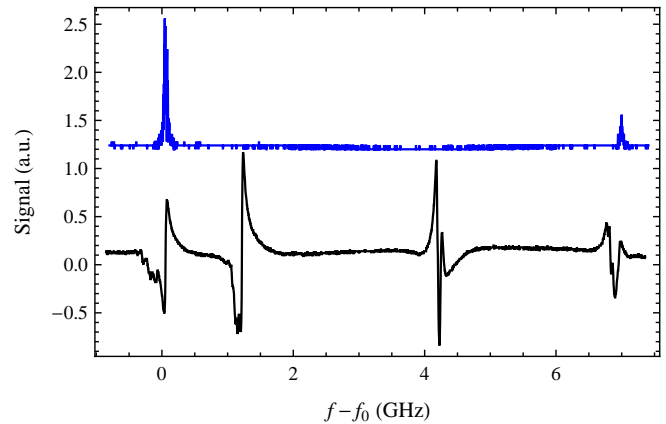}}
\caption{The spectrum of the slave laser with P1 oscillation (upper trace) and the D2 manifold polarisation spectrum of naturally abundant rubidium (lower trace). Only the positive frequency sideband is shown. The negative frequency sideband is typically barely visible above the noise level. The amplitude in the upper trace is approximately linearly proportional to power. }
\label{fig:heterodyne}
\end{figure}

\begin{figure}[htb]
  \centerline{\includegraphics[width=\linewidth]{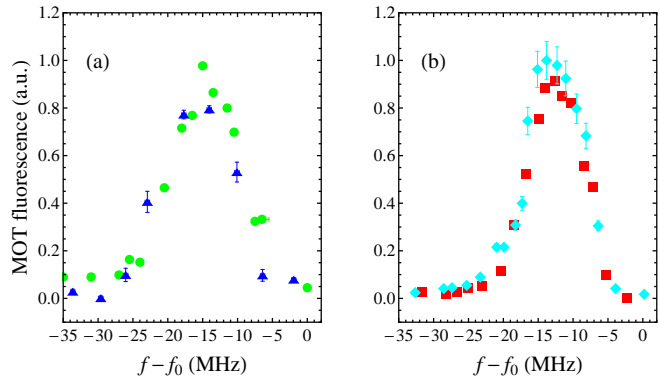}}
  \caption{MOT fluorescence with (a) unlocked slave laser with P1 oscillation (triangles) and (b) frequency locked laser with P1 oscillation (diamonds).  For comparison trapping light was generated by separate trap and re-pump lasers (circles and squares). We normalise the P1 data by factors 1.167 in (a) and 1.176 in (b) to account for the power lost in the sidebands. We have also taken differences in total trap light power into account. We estimate the peak atom count to be 1$\times 10^8$ in the MOTs made with separate lasers in both (a) and (b).
  }
  \label{fig:fluorescence}
\end{figure}

We set $f_i$ and $k$ so that P1 oscillation with a frequency $f_s \simeq$  \unit{6.8}{\giga\hertz} forms in the slave laser. The output then contains optical frequency components separated by the ground state hyperfine splitting in $^{87}$Rb. As an aide to coarse tuning of $f_s$ we now sweep the reference laser and  simultaneously record the entire rubidium D2 manifold spectrum using the spectrometer, and the slave laser spectrum using the heterodyne signal. The results are shown in  Fig.~\ref{fig:heterodyne} where the traces are offset for clarity.
A strong peak in the heterodyne signal can be seen near the optical frequency $f_0$ of the cycling  $5\textrm{s} ^2\textrm{S}_{1/2} (F=2) \to 5\textrm{p} ^2\textrm{P}_{3/2} (F'=3)$ transition of $^{87}$Rb, and a  weaker peak is evident at the
re-pumping $5\textrm{s} ^2\textrm{S}_{1/2} (F=1) \to 5\textrm{p} ^2\textrm{P}_{3/2} (F'=2)$ transition. We test the suitability of both an unlocked and a locked frequency P1 oscillation for generating MOT light. To make MOTs with  P1 oscillation on an unlocked slave, the master laser is stabilised to the re-pumping  transition. The slave laser is biased at 2.6 times the threshold current and we estimate that $f_r\simeq$ \unit{4}{\giga\hertz}. We set the injection parameters to $f_i$ = \unit{0.7}{\giga\hertz} and $k=0.052$.  While it is possible to tune $f_s$ by changing $k$ we find that in our setup which relies on mechanical control of the injection intensity, keeping $k$ fixed while adjusting the slave laser current (and thereby $f_i$)  provides better frequency control. Within about a \unit{3}{\giga\hertz} range, $f_s$ tunes approximately linearly with $\Delta f_s /\Delta f_i =$ \unit{0.64}{\giga\hertz\per\giga\hertz} and we use this sensitivity to finely tune the strong peak over the cycling transition. Fig.~\ref{fig:fluorescence}~(a) shows the resulting peak MOT fluorescence intensity during a dispenser pulse as a function of the detuning from line centre. The reference laser was locked to the $F=2 \to F'=2,3$ crossover feature at $f_0$-\unit{133}{\mega}{\hertz}, and the beat frequency was counted to provide a reference for the horizontal axis. For comparison, Fig.~\ref{fig:fluorescence}~(a) also shows data from MOTs made in the ordinary fashion with the slave laser stably injection locked without any sidebands and a separate re-pump laser.  

We also trap atoms with P1 oscillation on a frequency locked slave laser. In this case the master laser is stabilised to various points on the red side of the cycling transition, and $f_i$ = \unit{-11}{\giga\hertz} and $P_{inj}$ = \unit{6.9}{\milli\watt} which gives $k$ = \unit{0.14}.  This parameter combination leads to a weak sideband at the re-pump transition frequency. Now we find that $f_s$ is significantly less sensitive to variations in $f_i$ and we change $k$ to fine tune the sideband. The sensitivity of $f_s$ to $k$ is $\Delta f_s /\Delta P_{inj}=$ \unit{0.29}{\giga\hertz\per\milli\watt}. Fig.~\ref{fig:fluorescence}~(b) shows the peak fluorescence from MOTs with P1 oscillation on a frequency locked slave laser and again from MOTs made with a separate re-pumper for comparison. We have normalised the data to account for the power lost into the sidebands and changes in the coupling of light into the fibre were also taken into account. The error-bars on the ordinates are dominated by uncertainty in the normalisation factor. When trapping atoms with an unlocked slave, drifts in the free running slave laser frequency cause $f_s$ to drift by about  \unit{2}{\mega\hertz} during a dispenser pulse. This drift leads to the slightly larger uncertainty for the fluorescence count  on the slopes in Fig.~\ref{fig:fluorescence}~(a). For an injection locked oscillation the slave laser is stable to within \unit{1}{\mega\hertz}. The  re-pumping is relatively insensitive to fluctuations in $f_s$.
 
It can clearly be seen in Fig.~\ref{fig:fluorescence} that the location of the maximum and the overall shape of the peak fluorescence count does not depend on whether the light comes from one slave laser with P1 oscillation or separate trap and re-pump lasers. The data in Fig.~\ref{fig:fluorescence} (a) and (b) were taken with somewhat different MOT alignment which may be the cause of the small change in the width of the curves. Normalized temporal fluorescence data does not show noticeable differences in  MOT loading and lifetime when P1 oscillation light is used instead of separate lasers.  From this we conclude that aside from the power loss into the sidebands, P1 oscillation light traps atoms in a MOT just as efficiently as ordinary MOT light.  Our measurements show that the frequency stability of the P1 oscillation sideband is sufficiently good for MOTs even without active stabilisation of the injection parameters.  However, drifts during one day in the free running slave temperature causes $f_s$ of the unlocked P1 oscillation to drift enough that we need to adjust either $k$ or $f_i$ to recover the correct settings for atom trapping. In contrast, the injection locking band for the frequency locked P1 oscillation is large enough that the laser is unaffected by such drifts. Therefore we conclude that without active stabilisation the frequency locked P1 oscillation is a better choice for making rubidium MOTs. The long term stability of $f_s$ could be further improved upon by detecting the oscillation on a fast photodetector and stabilisation to a good microwave local oscillator either by controlling $k$ or $f_i$. 

We observe linewidth purity transfer from the master laser to the slave~\cite{Bondiou2000} in the short term spectrum of the beat-note between master and reference lasers. The spectral distribution is measured with a single trace on an RF spectrum analyser during around \unit{0.1}{\second} and the data is well modelled by Lorentzian fits. The linewidth measurements were carried out with the same laser settings as those used in the MOT experiment and the high bias current in the slave laser leads to a reasonably narrow full width at half maximum linewidth  of \unit{(3.0\pm0.3)}{\mega\hertz} in the free running slave. When optical injection produces P1 oscillation with an unlocked slave we find a linewidth of \unit{(1.1\pm0.2)}{\mega\hertz} for the low intensity feature at the optical injection frequency (i.e. the re-pumper in the MOT experiment) and \unit{(1.8\pm0.3)}{\mega\hertz} for the high intensity feature (the light which addresses the cycling transition). For the frequency locked P1 the corresponding linewidths were  \unit{(1.4\pm0.2)}{\mega\hertz} and \unit{(1.3\pm0.2)}{\mega\hertz}. The master-reference beat note has a Lorentzian linewidth of  \unit{(1.2\pm0.2)}{\mega\hertz}, which is likely to be dominated by the master laser since it has a much shorter extended cavity than the reference laser. In order to see a larger effect we injected into the slave laser with a much larger linewidth. We set the bias current in the slave laser to a lower value which produces a free running mode with a linewidth of approximately \unit{50}{\mega\hertz}. With injection parameters which yield qualitatively similar P1 oscillations to those produced with the higher bias current, the slave mode with lower bias becomes about \unit{1}{\mega\hertz} wide with injection. The P1 sidebands also become approximately as narrow. However, the slave laser is optimised to run at the higher bias by ensuring that the combination of bias current and temperature leads to a mode-hop free operation. At lower bias current and corresponding higher temperature we found that frequency jitter of the laser was too large and a Lorentzian model alone no longer fits all our data well. The narrowing effect is, however, very clearly visible.
 
The purity of the injected laser in the case of a P1 oscillation on a frequency locked slave laser can be understood as follows. With no injection the spontaneous emission into the lasing mode of the slave acts as a source for the laser. With higher injection intensity within the injection locking  frequency band the source becomes increasingly dominated by the master laser spectrum~\cite{Bondiou2000} which now acts as a phase reference. With P1 dynamics the relaxation oscillations are undamped and do not contribute significantly to the linewidth of the sideband. In the frequency locked case our experiment does not distinguish the linewidth of the spectral features of the P1 oscillation from the linewidth of the master laser mode. The reason we also observe linewidth narrowing for the large spectral feature in an unlocked slave with  P1 oscillation is less clear. The boundary between a bounded phase P1 and an unbounded phase P1 oscillation is not precisely known, and perhaps our injection parameters lead to dynamics where the master laser provides a partial phase reference. This hypothesis remains speculative and further studies of this intriguing phenomenon are clearly necessary. Nevertheless, for practical purposes under discussion here it is clear that the width of both the carrier and sideband features in the optical spectrum is well within the natual linewidth of many atomic transitions of interest. The linewidth of the P1 oscillation itself has been studied by direct measurement of the intensity oscillation with a high-speed photodetector~\cite{Chan2004}, and it was shown that the linewidth can be further reduced electronically to at least the instrumentation linewidth which was \unit{1}{\kilo\hertz} in that experiment. {This method leads to a high degree of coherence between the two features in the spectrum. The jitter in the absolute frequency is directly related to the jitter of the master laser which for most applications in atomic physics must be frequency stabilised.

By adjustment of the injection parameters we can create nearly single sided sidebands with an almost 1:1 sideband to carrier intensity ratio. 
The sideband intensity and frequency can be tuned independently by simultaneous adjustment of $f_i$ and $k$ using the maps in~\cite{Chan:07,eriksson2002} as a guide. The region in the $(f_i,k$) plane which produces a near 1:1 ratio in our slave laser is within the frequency locking range. We suggest that the P1 oscillation with this sideband intensity could be used to drive Raman transitions with applications in cooling~\cite{Kasevich1992} and coherent population trapping. With further amplification by a tapered amplifier, our technique can provide both the trapping and Raman cooling frequencies in cavity cooling schemes~\cite{Boca2004}. In this work we observe P1 frequencies which are five times larger than $f_r$. This measurement was limited by the tuneability of our lasers, and we believe much higher frequencies could be achieved. While we have only tested one particular model of laser we anticipate that due to the general nature of the nonlinear dynamics of optically injected diode lasers other models of diode lasers will show similarly large enhancements of bandwidth. The P1 sideband frequency
is in principle only limited by the free spectral range of the slave~\cite{Chan:07} which can easily be of the order of 100 GHz.  These frequencies far exceed those obtained by direct modulation of the diode laser current. Rotational frequencies in small heteronuclear molecules such as CaF~\cite{Wall2008} are in this range. Laser cooling of molecules could be extended to other species, and our technique offers a cheap way to simultaneously address two atomic or molecular transitions with tunable narrowband laser light provided resonant diode lasers can be found. We note that the slave laser emits the P1 sidebands with the same polarisation which makes delivery to the experiment very efficient.

Our sideband technique reduces MOT instrumentation, in particular for atomic species which either require an intense re-pumper or where the re-pump frequency lies beyond the modulation bandwidth. Since our technique only involves diode lasers miniaturisation is possible and since no active modulation is necessary the power consumption is low. Our sideband technique is attractive in combination with single beam scenarios such as the tetrahedron~\cite{Vangeleyn2009} and the micro-pyramid~\cite{trupke:071116}, and we hope it can further expedite the integration of light and atom traps on silicon chips~\cite{Pollock:09}.  With active frequency control our technique can be extended to applications which require a narrower linewidth than we achieve here. The  P1 oscillation on an unlocked slave  appears an ideal  source in experiments that require fast sweeps of one of the frequencies over large frequency bands. We anticipate that the bandwidth of the switching would be limited only by the dynamical frequency which is of the order of the P1 oscillation, and in principle very fast frequency switching is possible. 

In summary, we have used the P1 oscillation dynamics of an optically injected diode laser to create
MOTs with light derived from only the slave laser. This constitutes a simple but powerful demonstration of the applicability of the P1 oscillation sideband to all experiments in atomic physics which require two wavelengths with narrow linewdiths separated by the P1-oscillation frequency. We have discussed possible extensions of our work to larger P1 frequencies with applications in laser cooling of molecules.

\acknowledgments
The authors gratefully acknowledge Michael Charlton and Helmut Telle for the loan of equipment. We thank Dirk van der Werf and Hugh Thomas for help in setting up the laboratory and Julian Kivell for machining laser parts. This work is supported by the UK EPSRC and the Royal Society. 

\bibliographystyle{eplbib}

\end{document}